\PassOptionsToPackage{hyphens}{url}

\documentclass[%
 reprint,
 groupedaddress,
 showpacs, 
 noshowkeys,
 amsmath,amssymb,
 aps,
 prl,
]{revtex4-1}
\usepackage{color}
\usepackage{graphicx}
\usepackage{dcolumn}
\usepackage{bm}
\usepackage{amsmath}

\usepackage[cdot,mediumqspace,amssymb]{SIunits} 
\usepackage[autolanguage]{numprint} 

\usepackage{ifthen}
\usepackage{ifpdf}
\ifpdf
    \graphicspath{{Figures/PDF/}}
\else
    \graphicspath{{Figures/EPS/}}
\fi

\usepackage{breakurl}
\begin{document}

\title{Collective Decision Dynamics in the Presence of External Drivers}

\author{Danielle S. Bassett$^{1,2,*}$, David L. Alderson$^{3}$, Jean M. Carlson$^{1}$} \affiliation{$^1$Department of Physics, University of
California, Santa Barbara, CA 93106, USA; $^2$Sage Center for the Study of
the Mind, University of California, Santa Barbara, CA 93106; $^3$Naval
Postgraduate School, Monterey, CA 93943; $^{*}$ Corresponding author. Email
address: dbassett@physics.ucsb.edu}

\date{\today}

\begin{abstract}
We develop a sequence of models describing information transmission and
decision dynamics for a network of individual agents subject to multiple
sources of influence. Our general framework is set in the context of an
impending natural disaster, where individuals, represented by nodes on the
network, must decide whether or not to evacuate. Sources of influence include
a one-to-many externally driven global broadcast as well as pairwise
interactions, across links in the network, in which agents transmit either
continuous opinions or binary actions. We consider both uniform and variable
threshold rules on the individual opinion as baseline models for
decision-making.  Our results indicate that 1) social networks lead to
clustering and cohesive action among individuals, 2) binary information
introduces high temporal variability and stagnation, and 3) information
transmission over the network can either facilitate or hinder action
adoption, depending on the influence of the global broadcast relative to the
social network. Our framework highlights the essential role of local
interactions between agents in predicting collective behavior of the
population as a whole.
\end{abstract}

\maketitle

\section{Introduction}
The influences that humans have on one another's opinions and subsequent
actions can shape large-scale movements
\cite{Oliver1985,Simpson2004,Granovetter1973,Hedstrom2000}, and are often
facilitated by the sharing of information. Advances in information
technologies are rapidly changing the way that humans exchange and share
information. The widespread adoption of radio in the 1920s and television in
the 1950s ushered an era of live \textit{broadcast media}, greatly increasing
the speed and scope of mass communication from that of newspapers, which had
been the dominant form of ``one-to-many'' information dissemination since the
development of the Gutenberg printing press in the 15th Century. The
commercialization of the Internet in the late 1990s catalyzed the development
of new digital broadcast services, consumed by a diversity of computing
machines that now include laptop computers, mobile ``smart'' phones, and
other handheld devices. These devices are now ubiquitous; by mid 2010 there
were more than 5 billion mobile phone connections worldwide \cite{BBC2010},
with some regions experiencing more than 100\% penetration (meaning that
there is more than one mobile device per person).

A distinguishing feature of the Internet and these modern digital devices is
that they enable rapid ``one-to-many'' and ``many-to-many'' communication,
using services such as Facebook and Twitter, a phenomenon that has come to be
known as \textit{social media}. By the end of 2011 more than 300 million
users were accessing Facebook using mobile devices
\cite{BenEvans2011,HansenEtAl2011}. In this modern era, anyone with a digital
device and an Internet connection can publish information, and this is
dramatically changing the roles that individuals and corporations play in
traditional media industries such as books, music, film, and news journalism
\cite{KaplanMaenlein2010}.

There is a general recognition that these technologies allow information to
spread faster and perhaps more effectively. Such social epidemics, like
biological epidemics \cite{Dodds2005,Bettencourt2006}, can be thought of as
the result of single- \cite{Burt1987} or multi-stage \cite{Melnik2011}
complex contagion processes \cite{Centola2007,Centola2007b,Centola2010} that
can propagate through Facebook \cite{Onnela2010}, news websites
\cite{Leskovec2009} and other social media \cite{Simmons2011}, blogs
\cite{Leskovec2007} and Twitter \cite{Bakshy2011}.

Recent events on the East Coast of the United States have shown that social
media can be helpful during extreme weather events and natural disasters
\cite{Ball2011}. In the case of the 5.8 magnitude earthquake that occurred in
Virginia on 23 August 2011, the social network service Twitter proved to be
more reliable than cellular phones, which became overloaded by increased call
volume immediately after the event \cite{Conneally2011}, and news of the
event propagated to nearby New York on Twitter faster than the seismic waves
themselves \cite{Oswald2011}. Social media also allows for rapid organization
of disaster response information. For example, during the 7.1 magnitude
earthquake in the Canterbury region in the South Island of New Zealand on 4
September 2010, the University of Canterbury used Facebook and other
web-based technologies as a prominent source of up to date information and
support for many months \cite{Dabner2012}. It is less clear how this faster
information propagation affects the collective behavior of the individuals
who consume it. Social networking technologies have been credited with
facilitating the 2009 revolution in Moldova \cite{Synovitz2009} and the
``Arab Spring'' uprisings in 2011 \cite{HowardEtAl2011,KostTsvet2011}. More
prominently, social media has become an important component of corporate
marketing and advertising, with considerable effort now directed at
determining how to optimize the use of this new media
\cite{Frick2010,HannaEtAl2011,FastCompany2011}.

In this article, we consider the following question: {\em Do social
networking technologies like Facebook and Twitter ``help'' to bring a group
of individuals to action?} We consider the specific case of a population of
individuals, each of whom must decide if and when to commit to a binary
decision, and we assume that these individuals are exposed to information
both from broadcast sources and over social media. The study of information
diffusion on social networks has a lengthy history of illuminating the
large-scale spread of rumors, social norms, opinions, fads, and beliefs
\cite{Watts2002,Friedkin2011,Centola2007,Dodds2005,Easley2010,Valente1995,Rogers1995}.
The particular set of problems where individual decision-making occurs in the
presence of external information is known as ``binary decisions with
externalities'' \cite{Schelling1973,Watts2002,Salehi2006}. Here the
individual's decision is often modeled as a threshold on their underlying
opinion
\cite{Granovetter1978,Watts2002,Gleeson2007,Galstyan2007,Kerchove2009}, which
is modulated by the opinions of other individuals. We pose a minimal model of
individual decision-making and consider the effect of different forms of
information exchange on the behavior of the group as a whole.

\begin{figure}
\centering{}\includegraphics[width=.49\textwidth]{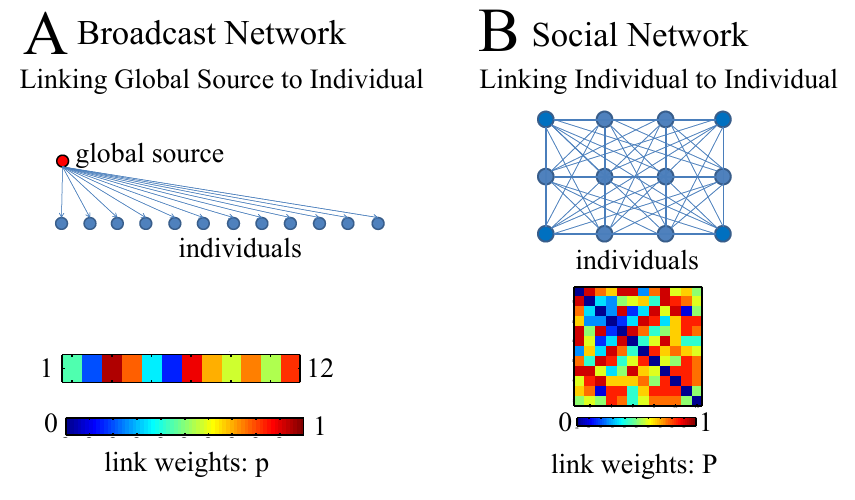}
\caption{\textbf{Model Construct} Here we illustrate our multi-layered
broadcast-social network system composed of a graph allowing information
diffusion between a global source and a population of 12 individuals
(broadcast network) and a second graph allowing information diffusion between
individuals (social network). \emph{(A)} The broadcast network structure is a
directed bipartite graph where the link weights are equal to the
probabilities that the global source transmits information to the individual
(probabilities given by the vector $p$, which varies over multiple numerical
simulations $M$). \emph{(B)} The social network structure is an all-to-all
undirected graph where the link weights are equal to the probabilities that
an individual transmits information to another individual (probabilities
given by the matrix $\mathbf{P}$, which varies over multiple numerical
simulations). In concrete terms, $P_{ij}$ can be thought of as the combined
rate that agent $i$ posts and agent $j$ reads the posting on Facebook.
\label{Fig0}}
\end{figure}

As a concrete example, we focus on a situation where a population is faced
with a pending natural disaster (e.g., hurricane or wildfire). In such
circumstances, it is know that each individual culls information from a wide
variety of digital, sensorimotor, and social sources \cite{Palen2007} and
must decide if and when to evacuate. We assume that each individual has a
simple \textit{decision rule}: if the individual believes that the disaster
is sufficiently likely, then he or she will evacuate. Individuals receive
information about the disaster from a ``global source'' that broadcasts
updates to the population as a whole (see Fig.~\ref{Fig0}A), and these
individuals also exchange information over a social network that allows them
to share opinions and observe the binary decisions of others (i.e., who has
evacuated; see Fig.\ref{Fig0}B). By numerically exercising a series of
increasingly complex models, we illustrate the tensions and trade-offs
inherent in social decision dynamics.

Our results suggest that information transmission over the social network can
either facilitate or hinder the action adoption depending on the influence of
the global source relative to the social network. Further, we find that the
sharing of binary information results in high variability, cascade-like
dynamics in which the time of collective action is difficult to predict.

\section{Model Construct}

We build a model of decision-making and social network interaction in
discrete time. The social networks we study can be thought of as Facebook- or
Twitter-like, in the sense that an agent posts updates or tweets to these
media and subsequently other agents can check the posting at some rate. In
our model framework, each agent receives a directed update from the other agents
one at a time at prescribed rates.

Let $t=0,1,2,\dots$ index the discrete time increments. The social network
consists of $n$ individuals, in which each agent $j=1,2,\dots,n$ has two
state variables:
\begin{itemize}
\item $S_{j}(t)$ = the \textit{internal state} at time $t$, where
    $S_{j}(t)\in[0,1]$;
\item $X_{j}(t)$ = the \textit{externally observable state} at time $t$,
    where $X_{j}(t)\in\{0,1\}$.
\end{itemize}
The internal state assumes continuous values, but the externally observable
state $X_{j}(t)$ reflects a binary \textit{decision} on the part of agent
$j$, derived from the \textit{decision rule}
\begin{equation}
X_{j}(t)=\left\{ \begin{array}{ll}
1 & \mbox{if }S_{j}(t)\geq\tau_{j}\\
0 & \mbox{if }S_{j}(t)<\tau_{j},
\end{array}\right.\label{eq:num_1}
\end{equation}
 and where $\tau_{j}\in[0,1]$ represents a \textit{threshold} value
for agent $j$. The use of a threshold for the binary decision $X_{j}(t)$ is
consistent with long-standing modeling efforts of collective behavior
beginning with the work of Granovetter in the 1970s \cite{Granovetter1978}
and is particularly relevant for decisions which are inherently costly
\cite{Watts2002} (such as evacuations).

We think of the internal state information as ``private'' in the sense that,
for example, it reflects an underlying belief on the part of the individual.
While it is possible that an individual could share this private information
with another, this exchange requires the individual to volunteer it. In
contrast, the externally observable state is ``public'' in the sense that
anyone who sees that individual would observe their binary decision. It has
been argued that this separation between internal opinion and external
decision is critical for an understanding of how convictions are coupled to
actions \cite{Martins2008,Martins2008b}.

We represent the internal state of all agents at time $t$ using
$S(t)=\{S_{1}(t),S_{2}(t),\dots,S_{n}(t)\}$, and we represent the externally
observable state of all agents at time $t$ using
$X(t)=\{X_{1}(t),X_{2}(t),\dots,X_{n}(t)\}$.

The internal state of an agent changes over time as a result of information
from three different sources: 1) a global source via a ``global broadcast'',
2) the internal state information of friends via ``social sharing'', and 3)
the decision state of neighbors via ``neighbor observation''. Each of these
sources is described in more detail below.

\paragraph*{Global Broadcast.}

We introduce a special external agent called the ``global source'' of
information that influences, but is not influenced by, the other agents. Let
$G(t)\in[0,1]$ represent the value that is broadcast by the global source in
time period $t$. We assume that receipt of a broadcast message by agent $j$
is binary (i.e., either it happens or it does not). Let $u_{j}(t)=1$
represent the successful broadcast from the global source to agent $j$ at
time $t$, and $u_{j}(t)=0$ represent no broadcast.
%
Thus the vector $U(t)=\{u_{1}(t),u_{2}(t),\dots,u_{n}(t)\}$
represents the overall broadcast from the global source at time $t$. This
global source is the primary ``external driver'' of dynamics in this
system.

\paragraph*{Social Sharing.}

We assume that the action of social sharing between agents is binary, either
occurring or not occurring at any point in time. That is, we let
$a_{ij}(t)=1$ if agent $i$ shares its internal state information with agent
$j$ in time period $t$, and $a_{ij}(t)=0$ if not. Thus, the matrix
$A(t)=\{a_{ij}(t)\}$ represents the adjacencies for exchange of internal
state information among the agents in time period $t$.
%
By construction, $a_{jj}(t)=0$ for all agents $j$ and all time periods $t$.
In general, we use symmetric sharing (i.e., $a_{ij}(t)=a_{ji}(t)$) to
indicate that any agent may address any other agent, however the mathematics
here do not require it. In our model, however, these links are directed in
the sense that for a given sharing event, an agent asks for another agent's
opinion or state, but does not share its own state. Whether or not there is
sharing symmetry or directionality depends on the specific application we
hope to model. Communication across cell phones tends to be symmetric and
undirected while that across Twitter and Facebook are not necessarily so.

\paragraph*{Neighbor Observation.}

Similarly, we assume that the observation of another agent's external state
either happens or does not in each time period $t$. Let $B(t)=\{b_{ij}(t)\}$
represent an adjacency matrix for exchange of externally observable binary
state information among the agents, where $b_{ij}\in\{0,1\}$. By
construction,  $b_{jj}(t)=0$ for all agents $j$ and all time periods $t$. The
matrix $B$ might represent a network of interactions based on physical
location, where nearby individuals literally see one another even if they do
not communicate. For purposes of our discussion, we refer to agents who are
adjacent in the matrix $B$ as \textit{neighbors}. That is, we narrowly define
two agents as ``friends'' if they share internal state information, and we
define two agents as ``neighbors'' if they can observe the external state of
each other. Again the mathematical formulation here can be used to model both
symmetric and non-symmetric sharing.


\paragraph*{General Update Rule.}

In the presence of all three sources of information, the internal state of
agent $j$ evolves according to the general update rule
\begin{equation}
\begin{split}
S_{j}(t+1) & = \\
& \frac{\displaystyle \sum_{i}a_{ij}(t)S_{i}(t)+ \displaystyle \sum_{i}b_{ij}(t)X_{i}(t)+u_{j}(t)G(t)}{{\displaystyle \sum_{i}a_{ij}(t)+\sum_{i}b_{ij}(t)+u_{j}(t)}}
.\label{update-eq}
\end{split}
\end{equation}
This update rule is a deterministic averaging of the current
internal state of agent $j$, the internal state of agent $j$'s friends, the
external state of agent $j$'s neighbors, and any global broadcast
information.
However, we assume that the coefficients in $A$, $B$, and $U$ are stochastic
in time;
that is, in any given time period $t$ the information to agent $j$
might or might not be received. This stochasticity is consistent with the
fact that communication over these types of networks is likely to be sparse
-- due to geographic and energetic constraints -- and dynamic -- due to agent
movement and constraints on communication \cite{Salehi2006}.

While opinion dynamics have been studied using a variety of models
\cite{Castellano2009}, the averaging rule given in Eq.~(\ref{update-eq}) is
most consistent with the Hegselmann-Krause model of opinion dynamics
\cite{Hegselmann2002}, models of ``continuous opinion dynamics''
\cite{Lorenz2005}, and the many models of coordination and consensus of
autonomous agents
\cite{DeGroot1974,JadbabaieEtAl2003,Salehi2006,SaberEtAl2007,MateiBaras2009}.
The discrete nature of this update rule is consistent with the fact that
information is often issued at some frequency or can be obtained in discrete
units from governmental, social, or technical sources
\cite{Drabek1986,Palen2007}.

\section{Collective Decision Dynamics}

In what follows, we are interested in the real-time exchange and mixing of
information among the individuals, and how it affects the decision-making of
the collective group.

We measure the behavior of the system as a whole in several ways. Recall that
the social network consists of $n$ individuals, in which each agent
$j=1,2,\dots,n$ has two state variables: $S_{j}(t)$ and $X_{j}(t)$. We refer
to the sequence $S(t),t\geq0$ which indexes over the $j$ agents as the
\textit{information state trajectory} for the system, and the sequence
$X(t),t\geq0$ similarly as the \textit{action adoption trajectory}. Let
$m=1,2,\dots,M$ index the numerical trials associated with a particular
experiment, and let $S_{j}^{m}(t)$ represent the value of $S_{j}(t)$ during
the $m^{th}$ trial. We compute the average information state of the
population at time $t$ during experiment $m$ as $\langle
S_{j}^{m}(t)\rangle_{n}=\frac{1}{n}\sum_{j=1}^{n}S_{j}^{m}(t)$; accordingly,
the sequence $\langle S_{j}^{m}(t)\rangle_{n},t\geq0$ is the \textit{average
information state trajectory} for the system during experiment $m$. We
compute the average information state of individual $j$ at time $t$ as
$\langle S_{j}^{m}(t)\rangle_{m}=\frac{1}{M}\sum_{m=1}^{M}S_{j}^{m}(t)$; this
is the average information state of any individual in the population.
Accordingly, the sequence $\langle S_{j}^{m}(t)\rangle_{m},t\geq0$ is the
\emph{ensemble information state trajectory} of a single individual over the
ensemble of numerical trials. Finally, we can estimate the \emph{average
ensemble information state trajectory }as
\begin{equation}
\begin{split}
\mathbb{E}_{m}\langle S_{j}^{m}(t)\rangle_{n} & =\frac{1}{M}\sum_{m=1}^{M}\left(\frac{1}{n}\sum_{j=1}^{n}S_{j}^{m}(t)\right) \\
& = \frac{1}{n}\sum_{j=1}^{n}\left(\frac{1}{M}\sum_{m=1}^{M}S_{j}^{m}(t)\right).\label{eq:ens_avg}
\end{split}
\end{equation}
It is important to note that the variance of $S_{j}(t)$ over individuals
could be very different from the variance expected over numerical trials; for
example, one distribution might be normal and the other heavy-tailed. The
separation of the individual and ensemble averages allows us to separately
probe these distinct sources of variability in the system.

We introduce the term $N(t)=\sum_{j=1}^{n}X_{j}(t)$, which counts the number
of individuals whose internal state has exceeded their decision threshold and
who therefore have taken action. With this definition, we observe that
$\langle X_{j}(t)\rangle=N(t)/n$ measures the fraction of individuals who
have chosen the binary action.

In many cases, we are also interested in the amount of time it takes for the
population to collectively adopt new information or take action. Define
$H_{S}(\alpha)=\min\{t\ |\ \langle S_{j}(t)\rangle\geq\alpha\}$ to be the
\textit{first hitting time} for the population to reach some \textit{average
information level} $\alpha\in[0,1]$. Similarly, define
$H_{X}(\alpha)=\min\{t\ |\ \langle X_{j}(t)\rangle\geq\alpha\}$ to be the
\textit{first hitting time} for the population to reach some \textit{average
adoption level} $\alpha\in[0,1]$. For example, $H_{X}(0.5)$ is the amount of
time for half the population to adopt the binary action.

Conceptually, we can think of two basic types of information diffusion; see
Fig.~\ref{Fig0}. The first is the (one-to-many) broadcast of external
information by the global source to the agents; we can think of this
interaction in terms of a broadcast network. The second (many-to-many)
diffusion occurs on two different social networks: one network of friends and
another network of neighbors. We study the behavior of the system in the
following specific cases.

\paragraph*{Case 1: Information broadcast from the global source only.}

In this case, the only information exchange is the broadcast from the global
source, and the update rule is given by
\begin{equation}
S_{j}(t+1)=\frac{S_{j}(t)+u_{j}(t)G(t)}{1+u_{j}(t)}.\label{u-update}
\end{equation}
In the simple
case where $S_{j}(0)=0$ and where the global source broadcasts $G(t)=1$ for
all time periods $t$, then the progression of internal state for agent $j$
goes as $1-1/(k+1)$ after the $k^{th}$ update. That is, with each update from
the global source, agent $j$ moves ``half way'' to the broadcast value of
$1$, reaching it only in the limit as $t\to\infty$.

As noted above, we assume that broadcast messages from the global source
are received stochastically.
This could occur in practice because agent $j$ only ``tunes in'' to the global
source sometimes or because the global source has limited success in its
ability to reach agent $j$.  For each time period $t$ we
generate a vector $U(t)$ as an independent and identically distributed random
vector, where $Prob\{u_{j}=1\}=p_{j}$, and this probability is independent
for each agent $j$. Thus, $\mathbb{E}[u_{j}(t)]=p_{j}$ captures the expected
broadcast ``rate'' \cite{Drabek1986} to agent $j$ (assumed to be stationary
for now), and $\mathbb{E}[U(t)]$ represents the overall expected broadcast
from the global source in time period $t$.

In this simple case, we can derive expected values for $S_j(t)$ and $X_j(t)$
analytically. Specifically, after $t$ discrete time units, the probability of
agent $j$ having received $k \leq t$ updates is given by the binomial
distribution, $\mathrm{Binom}(k; t, p_j)$.  The expected value of $S_j(t)$ is
therefore
\begin{equation} \label{ESjt}
\mathbb{E}[S_j(t)] = \sum_{k=0}^t  \binom{t}{k} (p_j)^k (1 - p_j)^{t-k} \left(1 - \frac{1}{k+1} \right).
\end{equation}
In addition, $\mathbb{E}[X_j(t)]$ is simply the probability that agent $j$
has taken action (i.e., has adopted state 1) and can be defined as
\begin{equation}
\begin{split} \label{EXjt}
\mathbb{E}[X_j(t)] & = Prob \{ S_j(t) \geq \tau_j \}\\
& = \sum_{k=0}^t
\binom{t}{k} (p_j)^k (1 - p_j)^{t-k} \ \mathbb{I}_{\{1 - \frac{1}{k+1} \geq
\tau_j \}},
\end{split}
\end{equation}
where $\mathbb{I}_{\{a\}}$ is the {\it indicator function}, namely
$\mathbb{I}_{\{a\}} = 1$ when condition $a$ is true, and $\mathbb{I}_{\{a\}}
= 0$ when condition $a$ is false.

The values $\mathbb{E}[S_j(t)]$ and $\mathbb{E}[X_j(t)]$ in the simple case
of global broadcast serve as a baseline against which we can evaluate the
impact of various types of social network exchange.

\paragraph*{Case 2: Global broadcast with social network exchange. }

We now add information exchange among friends alongside the global broadcast.
This gives us the following update rule:
\begin{equation}
S_{j}(t+1)=\frac{{\displaystyle \sum_{i}a_{ij}(t)S_{i}(t)+u_{j}(t)G(t)}}{{\displaystyle \sum_{i}a_{ij}(t)+u_{j}(t)}}.\label{caseii}
\end{equation}

Again, we assume that exchange of information between friends is stochastic
in time. Thus, we generate each matrix $A(t)$ as a weighted matrix, where
$Prob\{a_{ij}=1\}=P_{ij}$ and this probability is independent for each
$(i,j)$ pair. We let this probability be stationary in time, based on the
common assumption that the information transmission occurs on a faster time
scale than network changes \cite{Busch2012}. Thus,
$\mathbb{E}[a_{ij}(t)]=P_{ij}$ represents the expected rate at which agent
$i$ influences agent $j$, and $\mathbb{E}[A(t)]$ is the expected information
exchange within the social network of individuals, as might happen using
technologies such as mobile phones, Facebook, or Twitter.

This case corresponds to a particular type of consensus problem for which
there are analytic results that describe the convergence of
$\mathbb{E}[S_{j}(t)]$, specifically the conditions under which it is
guaranteed and how long it will take. Jadbabaie et
al.~\cite{JadbabaieEtAl2003} consider the case of ``leader following'' in
consensus problems in which one of the agents never updates its own variable,
but indirectly influences all of the other agents. The role of the ``leader''
is equivalent to that of our global source. (Bertsekas and
Tsitsiklis~\cite{BertTsik2007Comment} argue that this result is essentially a
special case of the more general result in \cite{TsitsiklisEtAl1986}.)
Similarly, Jadbabaie~\cite{Jadbabaie2004CDC} discusses routing in networks in
which nodes iteratively update their coordinate information, but certain
``boundary'' nodes retain fixed locations (again, like our global source). A
third related example is the model of Khan et al.~\cite{KhanEtAl2010}, which
describes consensus on random networks in which there are ``anchor nodes''
(again, like our global source in that their state does not change) and
``sensor nodes'' (like our agents who update through random mixing). Finally,
Galam et al.~\cite{Galam2007,Galam2010} use the term ``inflexible agents'' to
indicate nodes whose opinions do not change and find in some cases that these
agents drive opinion dynamics.

\paragraph*{Case 3: Global broadcast with binary information only among neighbors.}

In this final case, we consider the impact of binary exchange among neighbors
alongside the global broadcast. This gives us the following update rule:
\begin{equation}
S_{j}(t+1)=\frac{S_{j}(t)+{\displaystyle \sum_{i}b_{ij}(t)X_{i}(t)+u_{j}(t)}}{1+{\displaystyle \sum_{i}b_{ij}(t)+u_{j}(t)}},\label{caseiii}
\end{equation}

Again, we assume that this type of exchange is stochastic, and we generate
each matrix $B(t)$ as a weighted matrix, where $Prob\{b_{ij}=1\}=Q_{ij}$, and
this probability is stationary and independent for each $(i,j)$ pair. Thus,
$\mathbb{E}[b_{ij}(t)]=Q_{ij}$, and $\mathbb{E}[B(t)]$ is the expected
information exchange within local neighborhoods. In this case, we analyze the
network of friends separately from the network of neighbors. In future
simulations, it may be interesting to examine the special case in which the
two networks are either identical or change in a dependent manner.

The behavior of this system is sufficiently complicated that analytic results
do not, to our knowledge, exist for this case. We therefore turn to numerical
simulations in order to analyze and compare this case to the simpler ones
described above.

\section{Numerical Experiments}

We conduct several numerical experiments in which we simulate the behavior of
$n=100$ agents over a maximum of $T=1000$ time units.  Each agent $j$ starts
with $S_{j}(0)=0$, and the global source broadcasts $G(t)=1$ for all time
periods $t$.  For each of our three cases, we simulate a total of $M=100$
trials, and for each trial we select fixed broadcast rates for each agent
that are uniformly distributed on the interval [0,1] (that is, $p_j \sim
U[0,1]$).

Similarly, for each trial we select social sharing rates $P_{ij}$ or
neighbor observation rates $Q_{ij}$, each of which remain fixed within an
individual trial. We define a symmetric matrix $\mathbf{R}$ whose diagonals
decrease in value:
\begin{equation}
R_{i,i+|k|} = (n-k+1)/n ~ \forall ~k ~ \in \{1, 2, ... (n-1)\}.
\end{equation}
The matrix $\mathbf{R}$ can be thought of as a probabilistic form of a
regular lattice network. We choose $\mathbf{P}$ and $\mathbf{Q}$ as randomly
scrambled versions of $\mathbf{R}$, where the scrambling maintains the
symmetry of the matrix. In general, this approach provides flexibility for
the examination of topological structures between random and regular graphs.
In this work, we focus on random topologies to understand benchmark behavior.
The distribution of probabilities across agents provides a weighted
counterpart to degree heterogeneities examined in other studies of opinion
dynamics \cite{Lambiotte2007,Sood2005}, which have important consequences for
collective action.

For each of our three cases, we investigate the evolution of information
states and action adoption for individual agents and the collective group as
a whole.

\subsection{Average Information State}

\begin{figure*}
\centering{}\includegraphics[width=0.7\textwidth]{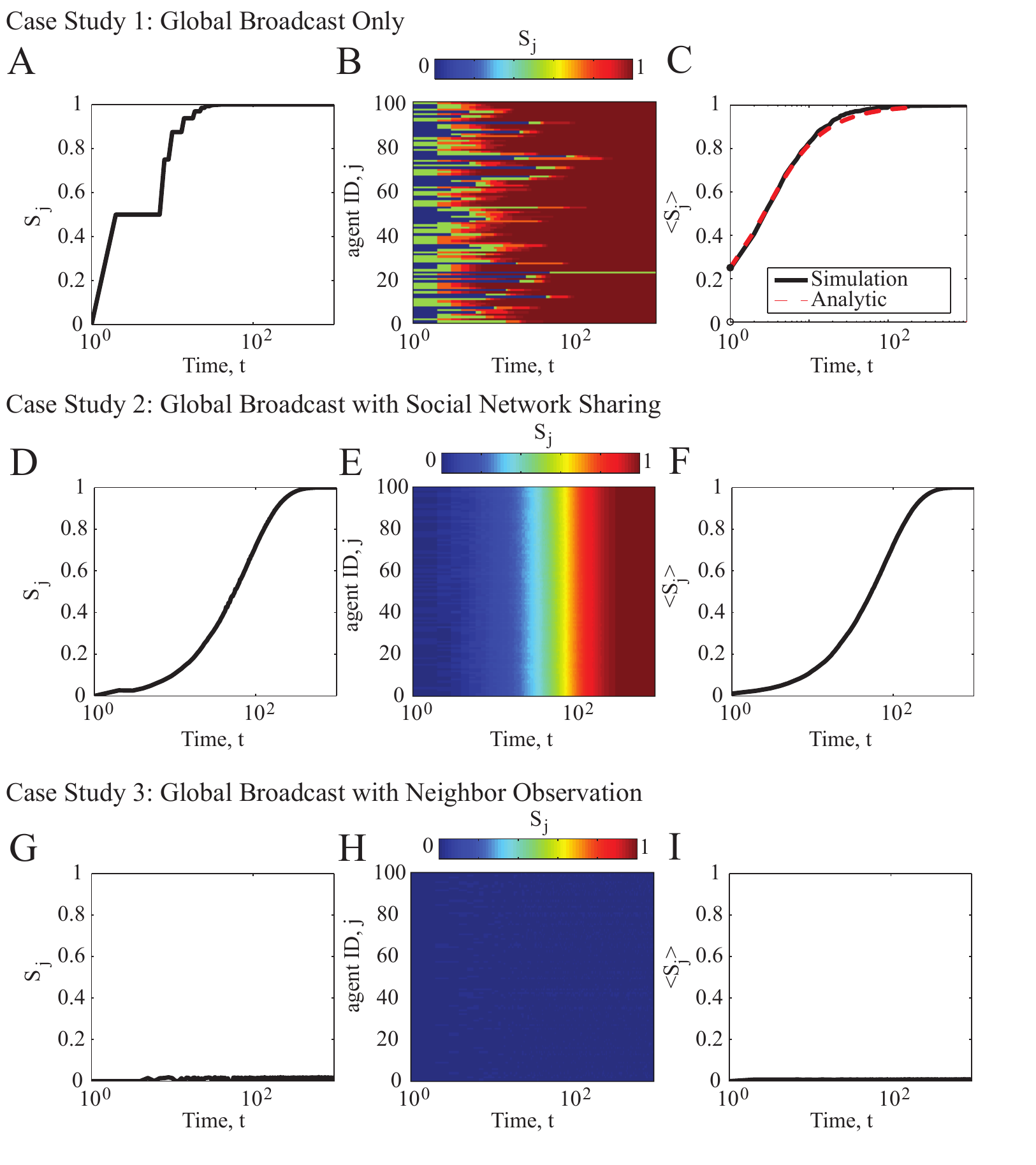}
 \caption{\textbf{Information Diffusion Under Three Types of Communication.}
 \emph{Panels A, D, G:} Information state trajectory for a single randomly selected agent $j$ as a function of time.
 \emph{Panels B, E, H:} Information state trajectories for a population of $n=100$
 agents. Information states (color) for each individual (row) are shown as a
 function of time (x-axis).
 \emph{Panels C, F, I:} Average information state variable, $\langle S_{j}\rangle$, averaged
over $M$=100 numerical simulations as a function of time. Note that the
simulated curve (black) in \emph{Panel C} is accompanied by a curve that was
analytically calculated from Eq.\ref{ESjt} (red).\label{Fig1}}
\end{figure*}

Figure \ref{Fig1} compares the evolution of information states across the
three case studies. Figure \ref{Fig1}A shows the stepwise trajectory $S_j(t)$
for a single agent, and Figure \ref{Fig1}B shows the overall population of
agents who independently update their state information in the presence of
the global broadcast only.  Figure \ref{Fig1}C presents the average
information state trajectory for the collective group calculated both through
simulation (black) and through the analytic solution given in Eq.
(\ref{ESjt}), which agree well.

Figures \ref{Fig1}D-F present the equivalent figures for the case of global
broadcast and social network sharing. We observe that in the presence of
social networking, individual agents (Figure \ref{Fig1}D) as well as the
entire group of agents evolve their state information in a collectively
smooth manner (Figure \ref{Fig1}E).  However, the overall rate of increase
for the average information state in the case of social network sharing
(Figure \ref{Fig1}F) is \emph{slower} than in the case of global broadcast
only (Figure \ref{Fig1}C).

Figures \ref{Fig1}G-I present the equivalent results in the case of neighbor
observation in which the only agent-to-agent information exchange comes from
observing the external state of other agents.  For the given parameters
($n$=100, $d$=1, $M=100$; for a description of $d$ see later sections), we
observe that none of the individuals ever raises their information state
significantly from the starting value of zero. This case represents an
extreme of the previous one in which social network sharing through neighbor
observation slows any rising of the information state. (Note: See later
sections for other areas of the state space in which potentially more
interesting behaviors like cascades and stagnation occur.)

\subsection{Average Adoption State}

\begin{figure*}
\centering{}\includegraphics[width=0.7\textwidth]{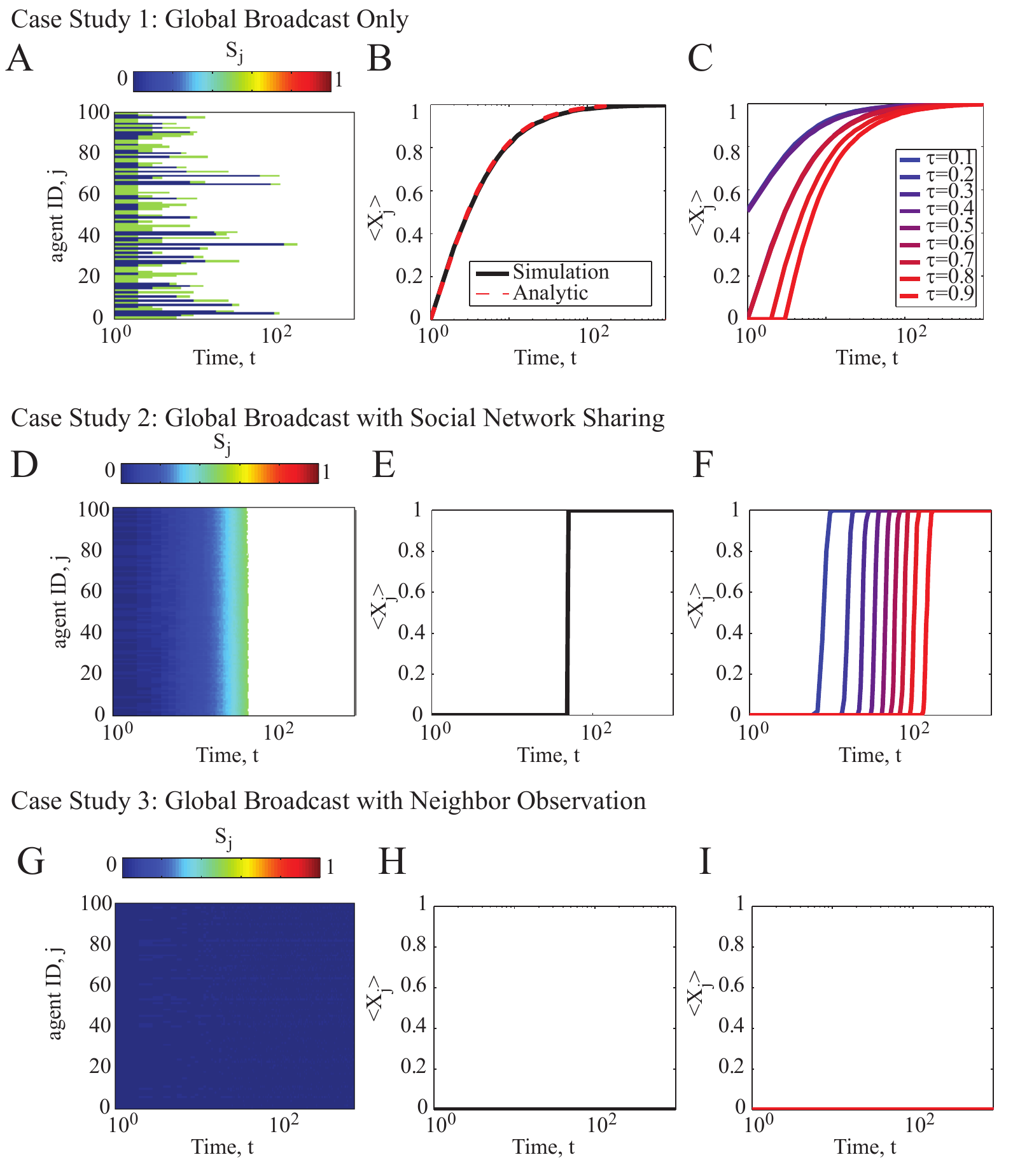}
\caption{\textbf{Action Adoption Under Three Types of Communication.}
\emph{Panels A, D, G:} Adoption state trajectories for a population of
$n=100$ agents. Adoption states (color) for each individual (row) are shown
as a function of time (x-axis). Note that the color white indicates that the
individual has taken action. \emph{Panels B, E, H:} Average adoption state
variable, $\langle X_{j}\rangle$, averaged over $M$=100 numerical simulations
as a function of time. \emph{Panels C, F, I:} Dependence of average adoption
state variable, $\langle X_{j}\rangle$, on the threshold level $\tau$, which
we have varied in this figure from 0.1 to 0.9. Note that the simulated curve
(black) in \emph{Panel B} is accompanied by a curve that was analytically
calculated from Eq. (\ref{EXjt}) (red). \label{Fig2}}
\end{figure*}

Figure~\ref{Fig2} displays the adoption trajectories for the three cases,
when each individual agent $j$ has a decision threshold $\tau_j = 0.5$.  In
Figure~\ref{Fig2}A, we observe that under only global broadcast the
information state of individual agents evolves in a manner consistent with
Figure~\ref{Fig1}B, except that we terminate each trajectory when the
individual crosses the decision threshold.  Figure~\ref{Fig2}B shows the
average adoption state trajectory for a single trial calculated both through
simulation and through the analytical solution given in Eq. (\ref{EXjt}),
which agree well. Importantly, to remain comparable to our simulations in
which $p$ is chosen $\sim U[0,1]$, Eq. (\ref{EXjt}) was calculated for a
range of $0<p_{j}<1$, and the average is plotted in Figure~\ref{Fig2}B. We
note that the action adoption curve for a population with fixed $p_j$ is not
equivalent to that for a population with distributed $p_{j}$. Finally,
Figure~\ref{Fig2}C shows the effect of different threshold values $\tau$ on
the average adoption state.

Figures \ref{Fig1}D-F present the equivalent results for the case of global
broadcast and social network sharing. As before, we observe that in the
presence of social networking, the entire group of agents evolves their state
information in a collectively smooth manner (Figure \ref{Fig1}D). Since this
groups shares a common decision threshold, we observe that action adoption
now occurs abruptly (Figure \ref{Fig1}E) and that the onset of this abrupt
change depends primarily on the threshold level $\tau$ (Figure \ref{Fig1}F).

In the case of neighbor observation and global broadcast, Figures
\ref{Fig2}G-I again show that none of the individuals ever raises their
information state significantly from the starting value of zero. These
results are consistent with those shown in Figure \ref{Fig1}G-I and represent
an extreme of social inertia keeping the information state at very low
values. (Again, see later sections for other areas of the state space in
which potentially more interesting behaviors like cascades and stagnation
occur.)

\subsection{Heterogeneous Decision Thresholds}

\begin{figure}
\centering{}\includegraphics[width=0.49\textwidth]{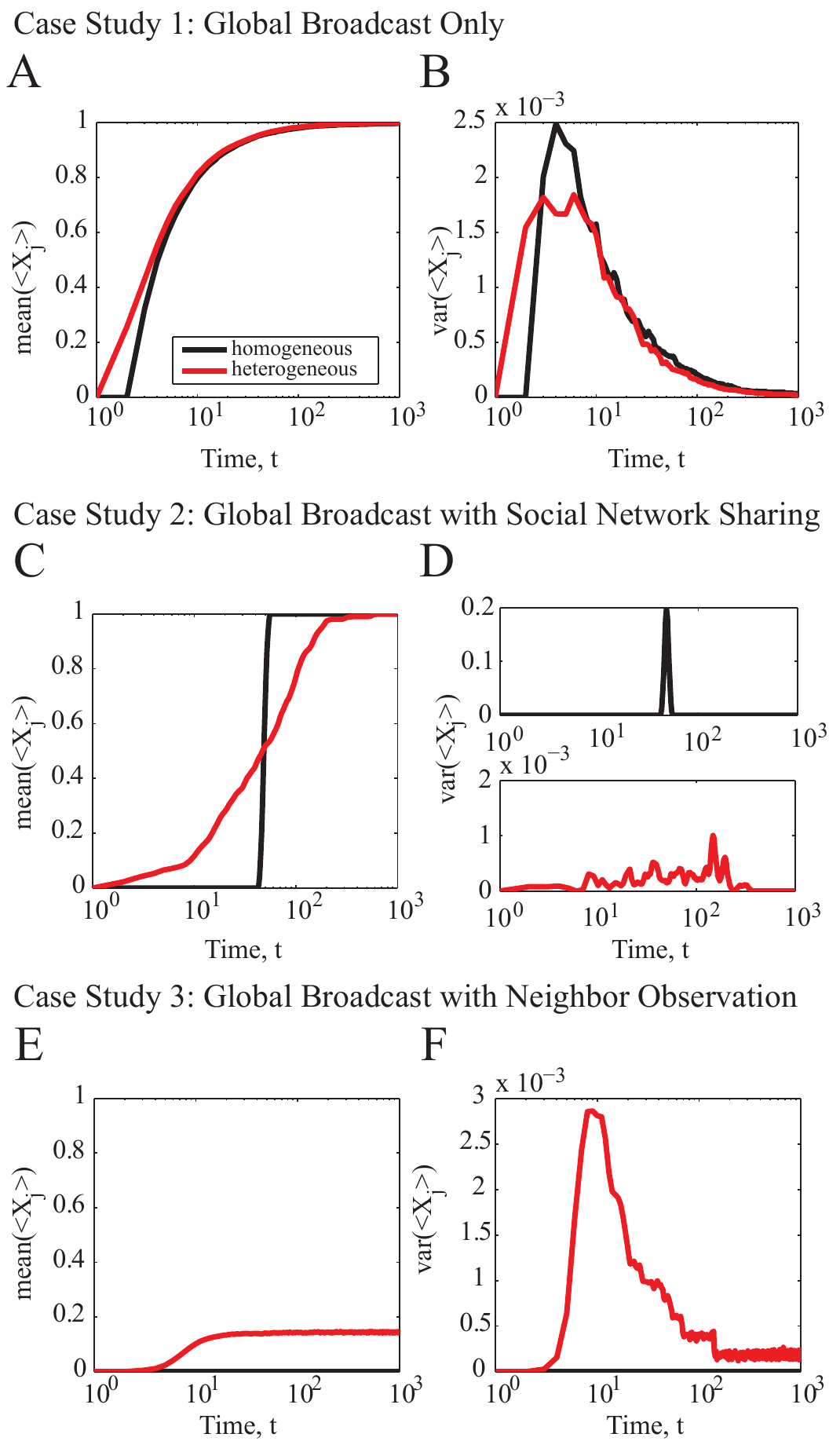}
\caption{\textbf{Action Adoption Under Heterogeneous Thresholds.} Thresholds
are either chosen to remain constant over all individuals (`homogeneous';
black lines) or to vary over individuals (`heterogeneous'; red lines).
\emph{Panels A, C, E:} Average adoption state variable, $\langle
X_{j}\rangle$, averaged over $M$=100 numerical simulations as a function of
time. \emph{Panels B, D, F:} The variance (over $M=100$ numerical
simulations) of the average adoption state variable, $\langle X_{j}\rangle$,
as a function of time. Note that the curves shown in panel (\emph{D}) are on
such different scales that we have plotted them in separate subplots to
enhance visualization. \label{Fig3}}
\end{figure}

Up to this point, we have assumed that individual agents share a common
decision threshold $\tau_j$.  In practice, this is unlikely to be the case
and recent theoretic and experimental work suggests individual variability in
decision thresholds may significantly alter the dynamics of the
population \cite{Bischi2009,Kearns2009}. We now consider the implications of a
heterogeneous population of agents where the threshold for agent $j$ is
uniformly distributed $\tau_j \sim U[0,1]$.  Figure~\ref{Fig3} displays the
impact of variation in underlying $\tau_j$ on the mean adoption state of the
collective group as a function of time when averaged over multiple trials,
given by (Eq. \ref{eq:ens_avg}), as well as the corresponding variance of the
adoption state of the collective group as a function of time.

We note that in the case of the global broadcast only, the distribution of
thresholds, whether homogeneous or heterogeneous, has little effect on the
mean or variance of the adoption state trajectories (Figure \ref{Fig3}A-B).

However, in the case of global broadcast and social network sharing, having a
homogeneous distribution of thresholds over the population leads to a
`tipping point', where at one time point no one has taken action while a few
time steps later, everyone has taken action (Figure \ref{Fig3}C, black line).
For heterogeneous thresholds, this drastic adoption is no longer evident, and
instead the transition from zero adoptions to all adoptions is smooth and
gradual (Figure \ref{Fig3}C, red line). Consistent with these results, we
find that the variance in the adoption states across individuals is large
precisely at the tipping point for homogeneous thresholds (Figure
\ref{Fig3}D, black line) and is small for all times when heterogeneous
thresholds are used (Figure \ref{Fig3}D, red line), suggestive of the
formation of a `collective'.

In the case of neighbor observation and global broadcast, Figure \ref{Fig3}E
shows that for this set of parameters ($\tau_j=0.5$ OR  $\tau_j \sim
U[0,1]$), when the threshold is distributed homogeneously the population is
unable to take action (black line) but when it is distributed
heterogeneously, the population can take action (red line). The fact that the
population can not take action when $\tau_j=0.5$ is true for all simulations,
leading to a zero-variance over simulations (Figure \ref{Fig3}F, black line).
In contrast, the action adoption trajectories for the population vary over
simulations when $\tau_j \sim U[0,1]$, and they do so particularly for time
points close to the point of maximum adoption (Figure \ref{Fig3}F, red line).

\subsection{Stagnation and Cascades}

The adoption state trajectories reported in the previous section were
averaged over $M$ numerical simulations. However, in order to fully
understand the case of neighbor observation and global broadcast, it is
important to examine individual simulation trials in addition to their
average.

\begin{figure}
\centering{}\includegraphics[width=.4\textwidth]{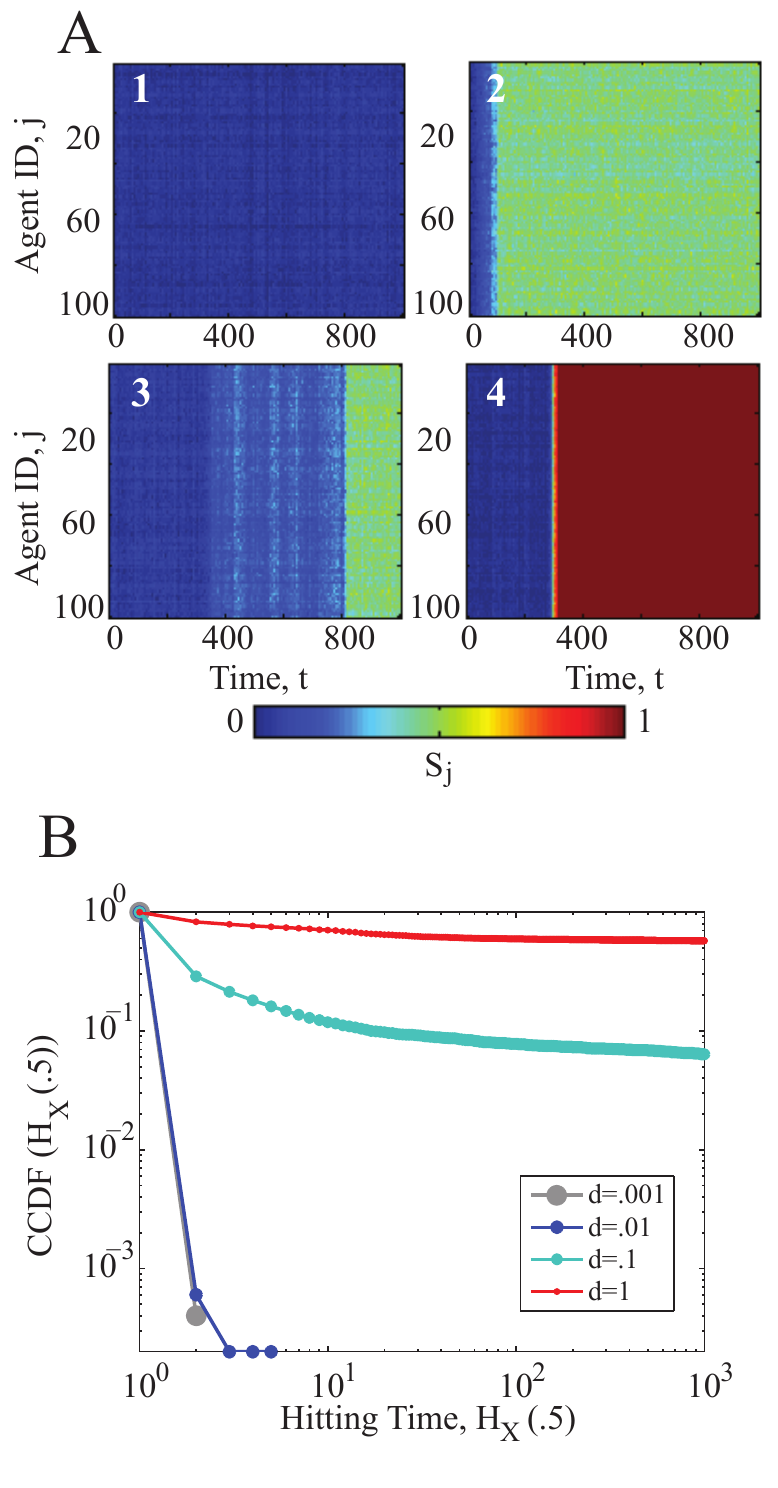}
\caption{\textbf{Variability in Action Adoption Behavior} (\emph{A}) Example
plot of the information state (color) trajectory of agents in the population
(y-axis) as a function of time (x-axis) showing (\emph{1}) stagnation,
(\emph{2}) partial cascade and then stagnation, (\emph{3}) late partial
cascade, and (\emph{4}) full cascade. (\emph{B}) The complementary cumulative
distribution function (CCDF) of the first hitting time for the population to
reach the average adoption level of 0.5 (e.g., $H_{X}(0.5)$) for a set of
$M=5000$ simulations as a function of the influence parameter $d$.
\label{Fig34}}
\end{figure}

In Figure \ref{Fig34}A, we show information state trajectories for all
individuals in a population in four different simulated trials. These four
examples highlight the different possible group behaviors including
stagnation (Figure \ref{Fig34}A1), partial cascades occurring either early
(Figure \ref{Fig34}A2) or late (Figure \ref{Fig34}A3), and full cascades
(Figure \ref{Fig34}A4). This high variability is an important factor that
sets Case 3 (neighbor observation and global broadcast) apart from the other
two cases. Cascade behavior has previously been demonstrated in wide variety
of topological structures and update rules
\cite{Watts2002,Gleeson2008,Gleeson2008b,Kerchove2009,Ghosh11wsdm} and
understanding when and how they occur if of particular import for predicting
large-scale social movements.

We can study this behavior more systematically by examining the first hitting
time for the population to reach an average adoption level of 0.5 (e.g.,
$H_{X}(0.5)$). We define a `converging simulation' as one for which
$H_{X}(0.5)$ is identified within $1000$ time steps and a nonconverging
simulation as one for which it is not. In Figure \ref{Fig34}B, we show the
complementary cumulative distribution functions (CCDFs) of converging
simulations for this model (\emph{red line}). We note that the majority of
converging simulations reach the target average adoption level in relatively
short times. However, the distribution is heavy-tailed such that much longer
times are consistent with the statistics. In fact, out of a total of $5000$
simulations, only $2127$ solutions converged in $1000$ time steps indicating
that stagnating periods can extend beyond the studied temporal window
($0<t<1000$). In the next section, we will further explore this behavior as a
function of the amount of exchange between social and global information
sources.

\subsection{Impact of exchange rates}

The behavior of the system under all three cases highlights the role of
different types of information exchange.  We observe that our stylized form
of social network exchange tends to move the group as a whole.

\begin{figure*}
\centering{}\includegraphics[width=.85\textwidth]{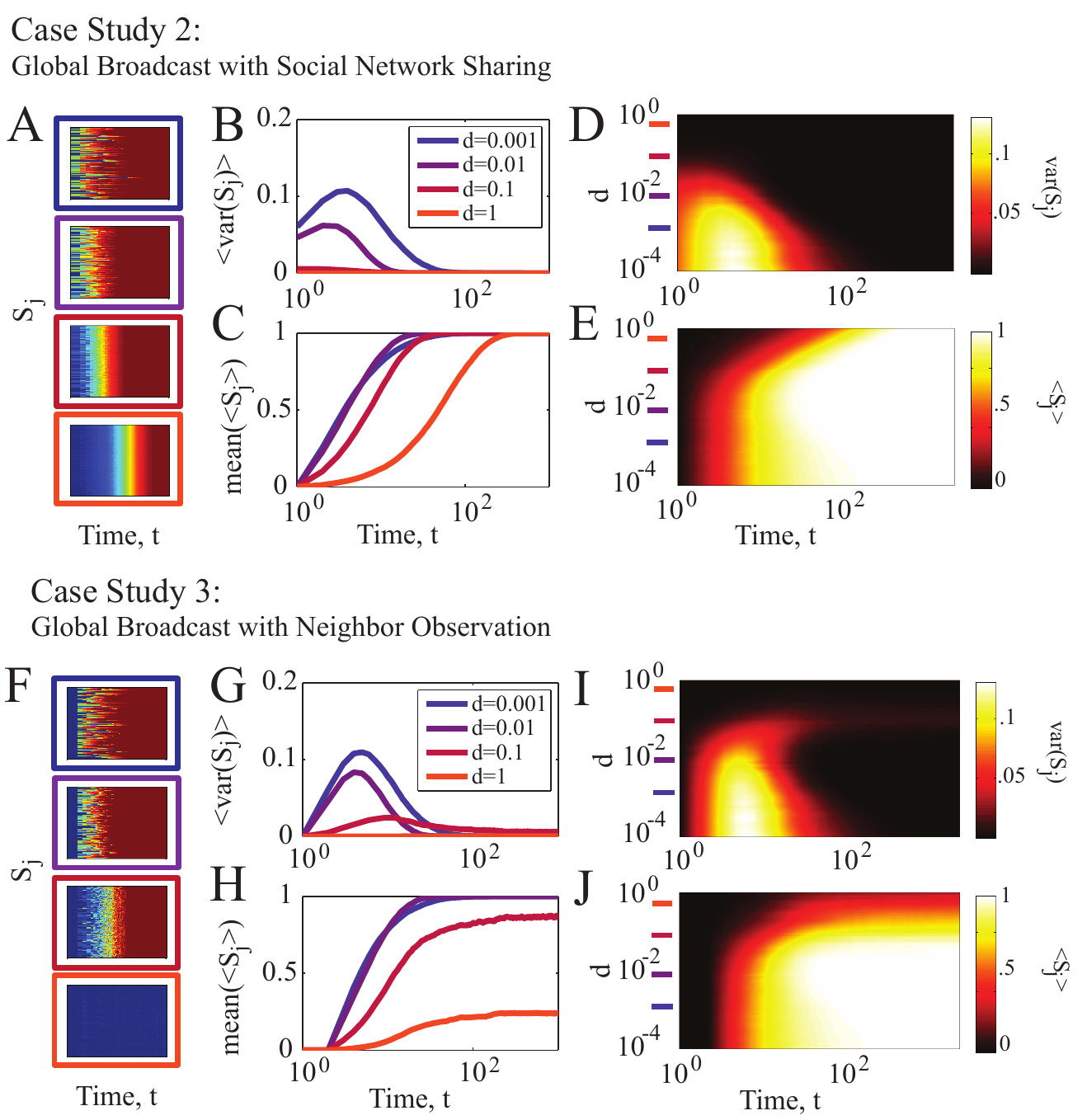}
\caption{\textbf{Effect of the Influence Parameter} in Case 2 (Global
Broadcast with Social Network Sharing; A-E) and Case 3 (Global Broadcast with
Neighbor Observation; F-J). (A,F) Information states for a population of
n=100 individuals as a function of time for four values of the influence
parameter: box outlines correspond to d-values given in the legends of panels
(B) and (G). As the influence parameter increases, the variance of
information states across individuals decreases (B, G), indicating the
formation of a 'collective', and the mean of information states across
individuals decreases (C, H), indicating that it takes more time for
individuals to reach higher values of $S_j$. Importantly, the four lines
shown in (B-C,G-H) are average (over $M=100$ simulations) trajectories drawn
from a continuous measure of influence, $d$. In (D-E, I-J) we therefore show
the full average contour plots for 7400 numerical simulations. Four colored
bars on the left of each plot indicate the d-values for which the four lines
in (B-C,G-H)  are taken. In these contour plots, color indicates the variance
(top) or mean (bottom) of the information state trajectory as a function of
time (x-axis) for influence parameters in the range $0.0001<d<1$ (y-axis).
\label{Fig4}}
\end{figure*}

In this section, we introduce an {\it influence} parameter $0 \leq d \leq 1$,
which serves to tune the network effects in Case 2 (social sharing) and Case
3 (neighbor observation), respectively, as follows:
\begin{equation} \label{dampedsocial}
S_{j}(t+1,d)=\frac{S_{j}(t)+{\displaystyle d \sum_{i \neq j} a_{ij}(t)S_{i}(t)+u_{j}(t)}}{1+{\displaystyle d \sum_{i \neq j} a_{ij}(t)+u_{j}(t)}}
\end{equation}
and
\begin{equation} \label{dampedneighbor}
S_{j}(t+1,d)=\frac{S_{j}(t)+{\displaystyle d \sum_{i \neq j} b_{ij}(t)X_{i}(t)+u_{j}(t)}}{1+{\displaystyle d \sum_{i \neq j} b_{ij}(t)+u_{j}(t)}}.
\end{equation}
When $d=1$ these cases remain unchanged and the social network has full
influence, but when $d=0$ these cases each reduce to global broadcast only
where the social network has no influence. This is illustrated in Figure
\ref{Fig4}A and F, where we see that for low values of $d$, the information
state trajectories are similar to those found in Case 1 (global broadcast
only; compare to Figure \ref{Fig2}B). The concept of network influence is
akin to the concept of information credibility, which varies over modern
information transmission technologies \cite{Flanagin2000,Metzger2008}: a
network with high credibility could be modeled as one with greater influence
and \textit{visa versa}.

As $d$ increases, we find that the variance in information state trajectories
across agents decreases (Figures \ref{Fig4}B and G), which is consistent with
the formation of a single `collective' state as the social network becomes
stronger. In addition, as $d$ increases we also find that it takes longer for
agents to reach any given state value, indicating that the social network is
maintaining some inertia and holding agents closer and closer to their
original states (Figures \ref{Fig4}C and H). These general behaviors are also
evident in a more continuous phase diagram of the state space of the system
(Figures \ref{Fig4}D-E and I-J). We note that the rightward tail of
nontrivial behavior in Figure \ref{Fig4}I in comparison to Figure
\ref{Fig4}D, and the extension of low $\langle S_{j} \rangle$ to longer times
in Figure \ref{Fig4}J in comparison to Figure \ref{Fig4}E are a result of the
stagnation followed by cascading behavior present for the case of neighbor
observation. These results indicate a wider range of active change or
variability in model behavior relative to the case of social network sharing.

\begin{figure*}
\centering{}\includegraphics[width=.85\textwidth]{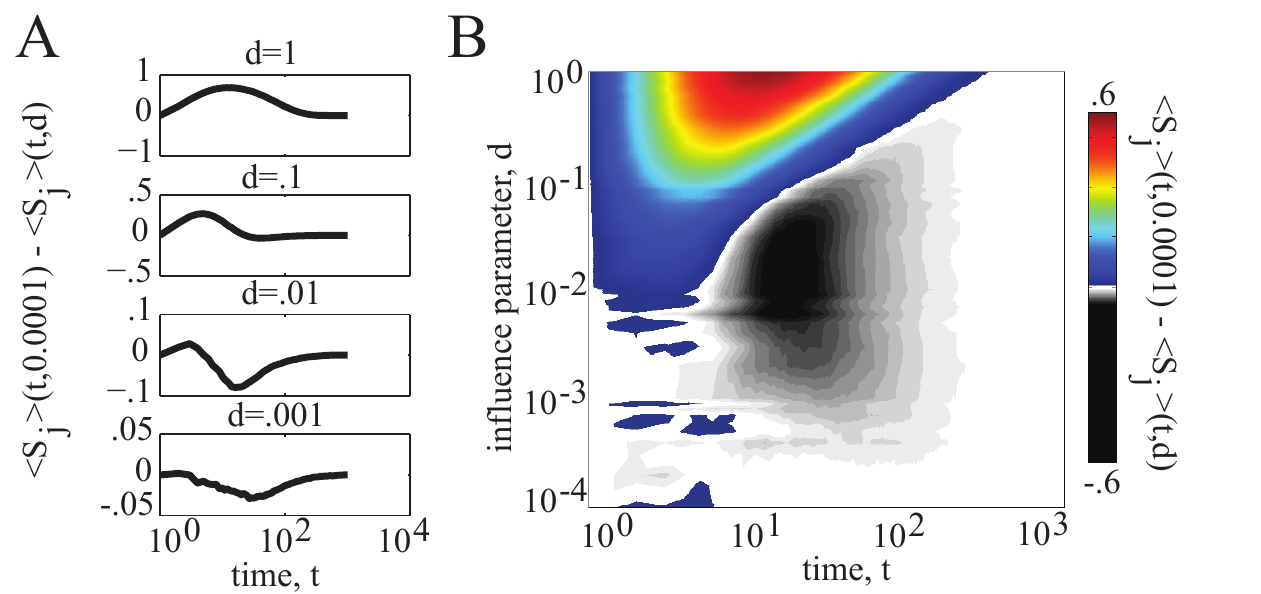}
\caption{\textbf{Facilitation and Hinderance by the Social Network} in Case
2. (\emph{A}) The difference between the average information state at
d=0.0001 ($\langle S_{j}(t,0.0001) \rangle$) and that at four other
increasing values of the influence parameter showing predominantly hinderance
($d=1$, \emph{top}; $d=0.1$ \emph{second row}), a combination of hinderance
and facilitation ($d=0.01$, \emph{third row}), and predominantly facilitation
($d=0.001$, \emph{fourth row}) by the social network exchange. (\emph{B}) The
larger surface from which the four curves in \emph{A} are drawn. Here we
compare $\langle S_{j}(t,0.0001) \rangle$ to $d$ ranging in $0.0001<d<1$.
Color indicates $\langle S_{j}(t,0.0001) \rangle - \langle S_{j}(t,d)
\rangle$. \label{Fig5}}
\end{figure*}

Importantly, the high variability in population behavior in Case 3 in the
previous section is present over a wide range of influence parameter values
(Figure \ref{Fig34}B). Here we study the first hitting time for the
population to reach an average adoption level of $0.5$ (e.g., $H_{X}(0.5)$,
when 50 out of a possible 100 agents have adopted). Accumulating data over
5000 numerical simulations, we find that it displays heavy tailed behavior:
in the majority of simulations, it takes a short time for those 50 agents to
take action, but in a few simulations, it takes a very long time.
Furthermore, this heavy-tailed behavior is sensitive to the influence
parameter $d$ and therefore the strength of the social network. For low
values of $d$, more numerical simulations have shorter first hitting times
which is consistent with what we would expect for situations that approximate
Case 1 (global broadcast only). This result is consistent with the fact that
the social network retards information state progress.

\subsection{Does the social network help or hinder?}

An important tradeoff is evident in Figure \ref{Fig4}C and \ref{Fig4}H for low values
of $d$. When $d=0.0001$, the average information state trajectory curve rises
the fastest initially but then slows for later values of $t$. In fact, for a
slightly larger value of $d=0.001$, the average information state trajectory
curve lags the $d=0.0001$ initially and then surpasses it later. This result
suggests that the social network -- if weakly present -- can help the entire
population take action sooner. However, when the social network becomes
stronger (e.g., $d=0.01$ and $d=1$), it acts like a cage, keeping information
states of all individuals from rising swiftly.

We investigate this behavior more systematically in Figure \ref{Fig5}, where
we observe three distinct behaviors: 1) facilitation of information
propagation for small rates of social exchange (e.g., small values of the
influence parameter $d$; Figure \ref{Fig5}A, bottom), 2) hinderance of
information propagation for large rates of social exchange (e.g., large
values of $d$; Figure \ref{Fig5}A, top), and 3) a combination of first
hinderance and then facilitation for intermediate rates of social exchange
(e.g., intermediate values of $d$; Figure \ref{Fig5}A, middle). Transitions
between these three distinct behaviors are smooth, as demonstrated in Figure
\ref{Fig5}B.

These results demonstrate that the social network can both hinder and
facilitate information state changes and by extension action adoption. When
the influence of the social network is large, the inertia of popular opinion
dampens the effect of the global source attempting to inject new information
into the system. On the other hand, when the influence of the social network
is small, the added mixing facilitates the dissemination of the new
information provided by the global source.

\section{Discussion}

Our long-term objective is to develop a framework that enables predictive
modeling of collective decision dynamics in situations involving multiple
sources of information. Modern communication technologies and social
networking applications provide fast, global means of information
dissemination. The need to determine the impacts of these technologies on
individual decisions and by extension collective action make it essential to
understand the interplay between multiple information sources.

This paper lays the foundation of such a framework, by systematically
exploring a sequence of models that aim to capture tradeoffs and tensions
that arise when a global broadcast source competes with information
transmission between individual agents. Despite our necessarily simplified
scenario, we find that information transmission over the network can either
facilitate or hinder action adoption, depending on the relative influence of
the global and social information sources. In most situations the social
network acts overall as a damping force, homogenizing opinion states and
delaying action adoption in the population.

\subsection{Insights from Biology}

Biological insights for the model behavior can be obtained by drawing
comparisons between the multi-layer information system and animal herding
behavior \cite{Couzin2005,Dyer2009,Leonard2012}. The combination of the
averaging update rule and the decision threshold forms the mathematical
framework for \emph{herding} or \emph{social conformity}
\cite{Banerjee1992,Milgram1969} in the sense that an animal can act based on
inferences from the information of other animals. In humans, models of such
information diffusion processes are built on a long history of empirical work
in sociology known as \emph{diffusion of innovations}
\cite{Coleman1966,Rogers1995,Strang1998}.

For a wide range of animals including humans, group decisions to move (for
food or travel) often depend on social interactions among group members
\cite{Couzin2005}, only a few of whom have pertinent information (e.g, food
location or migration route). So-called `informed' individuals correspond to
our global broadcast, while `uninformed' individuals correspond to our agents
on the social network. Unlike work investigating individuals who have no
decision preference \cite{Leonard2012}, our update rule hard-codes the fact
that agents in our system have a preference for retaining decision states.
Such an inertia is consistent with the observation that individual beliefs
are continually evolving variables that depends both on past beliefs
\cite{Solis2009} and newly acquired information, and in particular become
less malleable as time passes \cite{McCaffrey2004,Dash2007}.

\subsection{Insights from Statistical Mechanics}

Physical insights for the model behavior may be obtained by drawing an
analogy with the non-equilibrium statistical mechanics of a spin system
\cite{Sole2012,Yeomans1992}. Our global broadcast plays the role of a uniform
external field, while the information and adoption state variables of each
agent can be thought of as continuous or binary spins on a directed lattice
\cite{Castellano2009}. The initial state corresponds to all spins being
initialized at zero, with the external field fixed at unity. The interaction
between individual spins and the field is sampled stochastically, leading to
a noisy dynamical transition from inaction to action (a global attractor).

Inclusion of the social network corresponds to pairwise, directed
interactions between spins on a random lattice
\cite{Blatt1996,Bengtsson1995,Reichardt2006} that compete with the external
field. Because the spins are initialized at zero (opposite the field), the
social network initially tends to hinder action adoption, and in some cases
prevents action adoption entirely. Here inclusion of the network homogenizes
the collective behavior because the interaction between spins described by
the update rule is intrinsically stabilizing, damping the opinions of
outliers back toward the collective.

A familiar characteristic from the statistical physics of spins on a lattice
that is not observed in our model is the separation of agents/spins into
spatially localized domains characterized by action or inaction
\cite{Hed2001,Domany2001}. Two potential contributing factors are the damping
effect of our update rule and the mixing effect of the random social network
lattice, both of which inhibit local propagation of injected information. Our
preliminary investigations suggest that the update rule is the larger
predictor of behavior and therefore we expect that within our general
framework, clustering would occur for different broadcast and opinion update
rules. Exploiting parallels with well-understood systems in non-equilibrium
statistical physics, operations research, and graph theory are likely to
provide pathways for systematically unraveling the role of underlying network
structure, communication, and influence of collective behavior of
populations.

\subsection{Future Directions}

\paragraph{Update Rule} Our focus here is on the collective impact of individual decision-making
for when to evacuate, rather than the specifics of transportation and
routing. We implemented an unbiased rule for opinion updates, in which the
weight assigned to state variables is independent of the state value and the
time since the last update. This deliberately avoids destabilizing mechanisms
promoting microscopic propagation that arise in other contexts, such as the
spread of infectious diseases \cite{Barthelemy2005}.  In the case of
evacuations, it is plausible that recent information may be weighted more
heavily or travel preferentially along specific paths
\cite{Grabowicz2012,Zhao2010} in opinion updates, or an agent with new
information may be more likely to share information on the social network.
However, inclusion of these effects requires more sophisticated assumptions
about the individual agents that must be justified with cognitive and
behavioral data. While extracting influence and decision rules from network
databases remains a challenging problem in model identification, studies in
behavioral psychology, economics, and risk \cite{Glimcher2003,Glimcher2010}
may provide useful insights for more realistic representations of how
individual opinions are updated and decisions are made in the context of
social networks.

\paragraph{Network Topology} In this initial investigation, our model construct is deliberately chosen
to be generic, abstract, and random, setting a baseline for future work.  The
global broadcast is accessed at random by individuals, and the topology of
the social network is random as well.  In most situations the social network
acts overall as a damping force, homogenizing opinion states and delaying
action adoption in the community, a behavior that has also been identified in
experimental studies \cite{Weenig1991}. However, even in this scenario, there
are situations in which the social network accelerates action adoption.

In each case, design and optimization could naturally play a role in policy
decisions for specific scenarios.  For example, the topology of the social
network, both in terms of connectivity and rate of information flow, could be
based on realistic measurements of network traffic as measured by cellular
communication \cite{Gonzalez2008}, Twitter \cite{Weng2012}, or Facebook
\cite{Traud2012}. Alternatively, the social network for transmission of
information pertaining to action adoption during an evacuation might be
chosen to correspond to the geospatial layout of neighborhoods in a community
\cite{Onnela2011,Lambiotte2008,Crandall2010,Lerman12www}.  For a given
topology, global broadcast rates and/or transmission to individuals could be
tailored to the connectivity, and optimized for effectiveness as a guideline
for policy. The clustering inherent in such models will likely decrease the
cascade effects seen in our data \cite{Easley2010,Morris2000}.

\paragraph{Threshold Decisions} Our use of a uniform or random threshold rule for action adoption at the
individual level is a traditional starting point used in the decision-making
literature \cite{Granovetter1978} to study collective population dynamics.
Other factors that may influence decision-making may incorporate time
dependent effect---is an individual more likely to adopt an action when the
rate of action adoption is increasing in the population as a whole? Moreover,
the coupling between multiple networks -- e.g., human decisions and
transportation systems used for evacuation -- can play a vital role in the
collective behavior \cite{David2012} (e.g., when road congestion prevents the
group from evacuating effectively). A fundamental question is whether these
and other feedbacks are a cause or an effect of the underlying decision
process.

\subsection{Concluding Remarks}

In a broader context, the systematic development of a framework for
understanding the impact of social networks on collective behavior
corresponds to the development of a non-equilibrium statistical mechanics for
the impact of policy on decision-making in populations. Historically, many
fields in the social and life sciences have taken a phenomenological approach
to estimating the impact of deliberate policy and other externalities on the
behavior of populations. In this context, the study of social networks
corresponds to an underlying statistical mechanics for the collective
behavior, and must be understood systematically to obtain predictive behavior
of the population as well as a characterization of the variability within the
population.

It is increasingly recognized, across a broad range of fields, that
understanding network phenomena is essential to characterizing behavior of
the system as a whole. In the specific context of social systems,
interactions between individuals can for example give rise to financial
crashes \cite{Devenow1996}, political revolutions \cite{Kuran1989},
successful technologies \cite{Venkatesh2000}, and cultural market sensations
\cite{Salganik2006}. Constructing a statistical mechanics for these problems
is only a starting point. Feedback, design, optimization, and robustness are
all critical ingredients, mandating an interdisciplinary approach to
developing a reliable, predictive framework that is useful for policy.
Policy issues related to individual decisions including the impact
of training, identification of leaders, and signatures of stress and fatigue,
will be important topics of future research.

%

\paragraph{Acknowledgments} We thank Nada Petrovic, Emily Craparo, Mason A. Porter, Serge Galam, and
Andre\'{e} Martins for useful discussions. This work was supported by the
Errett Fisher Foundation, Templeton Foundation, David and Lucile Packard
Foundation, PHS Grant NS44393, Sage Center for the Study of the Mind, Office
of Naval Research MURI grants N000140810747 and 0001408WR20242, and NSF
(DMS-0645369).

\bibliographystyle{apsrev4-1} 
\bibliography{hd_bib}


\end{document}